# Optical and magneto-optical properties of ferromagnetic monolayer CrBr$_3$: A first-principles *GW* and *GW* plus Bethe–Salpeter equation study

Meng Wu, Zhenglu Li, and Steven G. Louie[*]

*Department of Physics, University of California at Berkeley, Berkeley, California 94720, USA*
*Materials Sciences Division, Lawrence Berkeley National Laboratory, Berkeley, California 94720, USA.*

[*]Email: sglouie@berkeley.edu

**Abstract**

The discovery of atomically thin two-dimensional (2D) magnetic semiconductors has triggered enormous research interest recently. In this work, we use first-principles many-body perturbation theory to study a prototypical 2D ferromagnetic semiconductor, monolayer chromium tribromide (CrBr$_3$). With broken time-reversal symmetry, spin-orbit coupling, and excitonic effects included through the full-spinor *GW* and *GW* plus Bethe–Salpeter equation (*GW*-BSE) methods, we compute the frequency-dependent layer polarizability tensor and dielectric function tensor that govern the optical and magneto-optical properties. In addition, we provide a detailed theoretical formalism for simulating magnetic circular dichroism, magneto-optical Kerr effect, and Faraday effect, demonstrating the approach with monolayer CrBr$_3$. Due to reduced dielectric screening in 2D and the localized nature of the Cr *d* orbitals, we find strong self-energy effects on the quasiparticle band structure of monolayer CrBr$_3$ that give a 3.8 eV indirect band gap. Also, excitonic effects dominate the low-energy optical and magneto-optical responses in monolayer CrBr$_3$ where a large exciton binding energy of 2.3 eV is found for the lowest bright exciton state with excitation energy at 1.5 eV. We further find that the magneto-optical signals demonstrate strong dependence on the excitation frequency and substrate refractive index. Our theoretical framework for modeling optical and magneto-optical effects could serve as a powerful theoretical tool for future study of optoelectronic and spintronics devices consisting of van der Waals 2D magnets.

I. Introduction

Chromium trihalides Cr$X_3$ ($X$ = Cl, Br, I) – magnetic semiconductors with layered structure in the bulk form [1–3] – have received enormous attention recently since the discovery of atomically thin 2D van der Waals magnets [4–10]. Atomically thin chromium trihalides exhibit high tunability with respect to electrostatic gating and magnetic fields [11–14], pressure [15, 16], and stacking order [17, 18], etc. Promising spintronics and valleytronics applications with van der Waals heterostructures consisting of atomically thin Cr$X_3$, such as the magnetic filtering effect [19, 20] and proximity effect [21, 22], have also been demonstrated in experiments.

Magneto-optical (MO) effects, including the MO Kerr effect, Faraday effect, and magnetic circular dichroism (MCD), are widely used as highly sensitive probes to characterize the electronic structure and magnetic properties of thin films [23, 24]. Among these, the MO Kerr effect has played an important role in the initial discovery of atomically thin intrinsic 2D magnets and demonstrated their rich magnetic behaviors [4, 5]. MO effects stem from the coupling between photons and the orbital motion of spin-polarized electrons, which is further interacting with the spin degree of freedom via spin-orbit coupling (SOC). Both the spin-splitting and SOC effects are essential for MO effects [25, 26]. In addition, the optical and MO properties of 2D magnetic semiconductors are strongly modified by the electron−hole interaction, forming tightly bound excitons that enhance the optical and MO responses [27–29]. To accurately model the optical and MO properties of 2D magnetic semiconductors, one needs to calculate the quasiparticle excitation energies including self-energy effects, and then calculate both the diagonal and off-diagonal elements of the dielectric function tensor including excitonic effects. These goals are achieved by using many-body perturbation theory such as the first-principles *GW* and *GW*-BSE methods [30–33], which have proven very successful in explaining and predicting the optical or MO properties of a variety of 2D materials of recent interest, such as monolayer transition metal dichalcogenides, black phosphorus, and CrI$_3$ [28, 34–38]. With our method, which supports full-spinor wavefunctions and broken time-reversal symmetry from the outset, we can achieve accurate quasiparticle band structure and dielectric responses from first principles, with which we can calculate the optical and MO properties.

So far, there have been several theoretical efforts using the *ab initio GW* and *GW*-BSE methods to study excited-state physics in atomically thin Cr$X_3$, such as the exciton-dominated optical and MO properties in monolayer CrI$_3$ [28], strong excitonic effects in the optical properties of monolayer, few-layer and bulk CrCl$_3$ [39], as well as chemical trends in the electronic structure, optical, and MO responses in monolayer Cr$X_3$ [40]. However, there is still ambiguity regarding whether self-energy effects on the quasiparticle band structure can be approximated through a rigid shift of the mean-field valence and conduction bands, as well as in how to calculate the MO signals from first principles including the important excitonic effects. In this paper, we use ferromagnetic monolayer CrBr$_3$ as a model system. As a member of the Cr$X_3$ family, monolayer CrBr$_3$, which is an air-stable magnetic semiconductor with non-negligible SOC effects, provides an ideal platform for exploring fundamental physics and for potential applications of spintronics devices [41, 42]. However, a detailed and thorough theoretical study of self-energy and excitonic effects has been lacking. To this end, we perform full-spinor *GW* and *GW*-BSE calculations of ferromagnetic monolayer CrBr$_3$ and demonstrate the importance of SOC, self-energy, and excitonic effects. We further lay out the formalism to simulate the optical and MO properties with the calculated *ab initio* frequency-dependent layer polarizability tensor and dielectric function tensor.

The rest of this paper is organized as follows. In Sec. II, we discuss the crystal structure of monolayer CrBr$_3$ and the importance of SOC effects. Computational methods are explained in detail. In Sec. III, we discuss the treatment of broken time-reversal symmetry in the *GW* method and present the computed quasiparticle band structure results. In Sec. IV, we calculate the exciton eigenstates, and construct both the diagonal and off-diagonal elements of the layer polarizability tensor as well as the dielectric function tensor. We simulate the absorbance spectrum with linearly polarized light, and analyze the exciton energy levels and exciton amplitudes in real space. In Sec. V, the formalism of simulating MO effects is presented. We then calculate the MCD of the absorbance spectrum. Finally, to connect to experiments, a two-interface polar setup with normal incidence is used to simulate the MO Kerr and Faraday effects in monolayer CrBr$_3$ with different substrates.

**II. Crystal structure and SOC effects**

Bulk CrBr$_3$, crystallized in the rhombohedral BiI$_3$ structure type <420 K, is a van der Waals layered material, with the space group $R\bar{3}$ [2, 43]. Below the Curie temperature of ~ 37 K, long-range ferromagnetic order emerges with an out-of-plane easy axis [44, 45]. Within each atomic layer, Cr atoms are arranged in a honeycomb structure, with each surrounded by six bromine atoms arranged in an octahedron, as shown in Figs. 1(a) and 1(b). In its monolayer form, CrBr$_3$ has a point group of $S_6$, and a Curie temperature ~27 – 34 K [41, 42, 46]. The octahedral crystal field splits the Cr $d$ orbitals into $t_{2g}$ and $e_g$ manifolds, of which the spin degeneracy is further broken by the magnetic exchange interaction, as shown in Fig. 1(c). Hybridization between Cr $d$ orbitals and Br $p$ orbitals in the octahedral crystal field forms ligand states and broadens the band dispersion [47] (see Sec. I of the Supplemental Material [48]). Because Cr is in the $3d^3$ electronic configuration, each Cr site hosts a magnetic moment of $3\mu_B$ (with fully occupied majority-spin $t_{2g}$ orbitals), according to the Hund's first rule. Recent *in situ* spin-polarized scanning tunneling microscopy and spectroscopy experiments have confirmed the monolayer crystal structure and the out-of-plane easy axis [17]. Throughout this paper, the magnetization of ferromagnetic monolayer CrBr$_3$ is taken to be along the +$z$ direction. The majority-spin polarization direction is denoted as spin-up.

In this paper, we used density-functional theory (DFT) and the method of local-spin-density approximation with an on-site Hubbard potential (LSDA+$U$) to serve as a reasonable mean-field starting point for the following $GW$ (at the $G_0W_0$ level) and $GW$-BSE studies. The LSDA+$U$ method employed is implemented in the Quantum ESPRESSO package [49, 50]. We took the on-site Hubbard interaction parameter $U$ = 1.5 eV and Hund's exchange interaction parameter $J$ = 0.5 eV [51], the validity of which have been justified in previous theoretical works of chromium trihalides, sharing similar crystal structures and chemical environment [28, 39, 40]. A supercell model with a thickness of 16 Å along the direction normal to the layer was adopted to avoid interactions between periodic images. We employed optimized norm-conserving Vanderbilt pseudopotentials including Cr $3s$ and $3p$ semicore states [52]. The Kohn-Sham orbitals were constructed with a plane-wave energy cutoff of 70 Ry. We used the experimental monolayer lattice constant $a$ = 6.3 Å [17] and relaxed the internal coordinates until the forces are converged within 5 meV/Å. The relaxed structure has a Cr-Br-Cr bond angle of 94.84°, consistent with the ferromagnetic super-exchange interaction [53]. In both DFT and $GW$ calculations, we truncated the Coulomb interaction in the $z$-direction as discussed in Refs. [54–56].

In the materials family of Cr$X_3$ ($X$ = Cl, Br, I), the SOC strength and magnetic anisotropy increase with the atomic mass of the halogen element [2, 42, 57]. On the one hand, previous first-principles studies on excitonic effects in monolayer and few-layer CrCl$_3$ have neglected SOC effects [39], which is justified by its small magnetic anisotropy [58]. On the other hand, monolayer CrI$_3$ hosts strong SOC strength and forms a highly anisotropic Ising-type spin system with an out-of-plane easy axis, which means a first-principles modeling of its optical and MO properties must include SOC effects [28, 40]. It is therefore interesting to first study SOC effects on the Kohn-Sham (mean-field) DFT band structure of monolayer CrBr$_3$, which shows reduced anisotropy compared with CrI$_3$ in the bulk and few-layer forms [42, 59]. In Figs. 1(d) and 1(e), we compare the band structure of monolayer CrBr$_3$ without and with SOC effects, respectively. Monolayer CrBr$_3$ has an indirect band gap of 1.67 eV both without and with SOC; the valence band maximum is at the *M* point while the conduction band minimum is along the Γ-*K* path. The direct band gap at the *M* point is 1.69 eV without SOC and 1.68 eV with SOC. Without SOC effects, spin is a good quantum number, and the band structure in Fig. 1(d) can be grouped into spin-up and spin-down bands. The energy ordering of Cr *d* states, as shown in the projected density of states (DOS) plots in Figs. 1(d) and 1(e), agrees well with our above analysis of the Cr *d*-orbits in an octahedral crystal field. When we switch on SOC effects, there are noticeable changes to the band structure, as shown in Fig. 1(e). First, two-fold degeneracies at the Γ point are lifted, because the double group $S_6^D$ (to which the Bloch states at the Γ point belong in the presence of SOC) has only one-dimensional irreducible representations. Second, spin is no longer a good quantum number and some lower-energy valence states have mixed spin polarization. Third, according to the projected DOS plot in Fig. 1(e), spin-down Cr *d* orbitals have more contribution to the top valence states when SOC effects are present. In this way, we conclude that SOC effects are important in describing the electronic structure of monolayer CrBr$_3$, and therefore using full-spinor wavefunctions in the *GW* and *GW*-BSE formalism is essential [28, 38].

**III. Quasiparticle band structure**

In this section, we describe the computational details of our *GW* calculations at the $G_0W_0$ level and compute the quasiparticle band structure of monolayer CrBr$_3$. An accurate first-principles modeling of the electronic structure of monolayer CrBr$_3$ needs to account for the magnetic order, the dielectric screening in a quasi-2D environment, as well as the on-site Coulomb interaction among the localized spin-polarized electrons. Through the screened Coulomb interaction *W*, the nonlocal and dynamical screening effects beyond DFT-LSDA can be captured accurately. In the presence of long-range magnetic order, there is no time-reversal symmetry in the system. We therefore use the full Adler–Wiser expression to calculate the static irreducible polarizability within the random-phase approximation (RPA) used in the Hybertsen–Louie plasmon-pole model [30, 60]:

$$\tilde{\chi}_{\mathbf{G}_1\mathbf{G}_2}(\mathbf{q}) = \frac{1}{N_k V} \sum_{cv\mathbf{k}} \left\{ \frac{\langle v(\mathbf{k}-\mathbf{q})|e^{-i(\mathbf{q}+\mathbf{G}_1)\cdot\mathbf{r}}|c\mathbf{k}\rangle\langle c\mathbf{k}|e^{i(\mathbf{q}+\mathbf{G}_2)\cdot\mathbf{r}}|v(\mathbf{k}-\mathbf{q})\rangle}{\epsilon^{\text{MF}}_{v(\mathbf{k}-\mathbf{q})} - \epsilon^{\text{MF}}_{c\mathbf{k}}} \right. \\ \left. + \frac{\langle c\mathbf{k}|e^{-i(\mathbf{q}+\mathbf{G}_1)\cdot\mathbf{r}}|v(\mathbf{k}+\mathbf{q})\rangle\langle v(\mathbf{k}+\mathbf{q})|e^{i(\mathbf{q}+\mathbf{G}_2)\cdot\mathbf{r}}|c\mathbf{k}\rangle}{\epsilon^{\text{MF}}_{v(\mathbf{k}+\mathbf{q})} - \epsilon^{\text{MF}}_{c\mathbf{k}}} \right\}, \quad (1)$$

where $|v(\mathbf{k}\pm\mathbf{q})\rangle$ denotes a valence band eigenstate with band index $v$ and crystal momentum $\mathbf{k}\pm\mathbf{q}$, $|c\mathbf{k}\rangle$ denotes a conduction band eigenstate with band index $c$ and crystal momentum $\mathbf{k}$, and $\epsilon^{\text{MF}}_{v(\mathbf{k}\pm\mathbf{q})}$ and $\epsilon^{\text{MF}}_{c\mathbf{k}}$ are the mean-field energies of corresponding states, respectively. Here, $\mathbf{G}_1$ and $\mathbf{G}_2$ refer to reciprocal lattice vectors which form the Bravais lattice in the reciprocal space. Also, $N_k$ is the number of $k$-points, and $V$ is the volume of a unit cell. The static irreducible polarizability $\tilde{\chi}$ is then used to calculate the screened Coulomb interaction $W$, as detailed in Ref. 30. Note that, here, the two terms in the summation cannot be combined into one using time-reversal symmetry. The full Adler–Wiser expression is crucial to keep the particle exchange symmetry of the screened Coulomb interaction such that $W(1,2) = W(2,1)$, where 1 and 2 each denote coordinates (space, spin, and time) of a particle. Since we choose LSDA+$U$ as our mean-field starting point for the subsequent $G_0W_0$ calculations, we treat on the same footing the Hubbard potential ($V_{\text{Hub}}$) and the LSDA exchange-correlation potential ($V_{\text{xc}}^{\text{LSDA}}$). That is, we subtract these mean-field contributions to the exchange-correlation potential in the $G_0W_0$ self-energy and use the difference as a first-order perturbation on the mean-field band energies [30]. The self-energy correction operator is given by [61–63]

$$\Delta\Sigma = \Sigma - V_{\text{xc}}^{\text{LSDA}} - V_{\text{Hub}}, \quad (2)$$

where $\Sigma$ is the self-energy operator in the *GW* approximation [64].

The *GW* (at the $G_0W_0$ level) and *GW*-BSE calculations – for the quasiparticle band structure and optical properties, respectively – were performed using the BerkeleyGW package [55]. Since the standard sum-over-bands approach [30] and the static remainder approach [65] in calculating the *GW* self-energy converge to the final result in opposite directions, we took their average to speed up the convergence of quasiparticle band energies with respect to the number of empty states. The kinetic energy cutoff in calculating the screened Coulomb interaction was set to 50 Ry, and a total of 2,000 bands were included in the *GW* calculation, with the highest band energy at ~90 eV above the valence band maximum. We adopted a $3 \times 3 \times 1$ *q*-grid with three subsampling points using the nonuniform neck subsampling method to describe the 2D dielectric screening [66]. The quasiparticle band gap was converged to within 50 meV. To perform Wannierization of the Cr 3*d* orbitals and Br 4*p* orbitals (a total of 56 orbitals) for later interpolation of the quasiparticle band structure, the self-energy corrections of 42 valence bands and 14 conduction bands were calculated on a $33 \times 33 \times 1$ grid. We treated the dynamical screening effect through the Hybertsen-Louie plasmon-pole model [30], where the charge density of the itinerant valence states (42 in total, excluding semicore Cr 3*s* and 3*p* states) were used. The resulting quasiparticle band structure was interpolated with spinor Wannier functions, using the Wannier90 package [67].

In Fig. 2, we show the quasiparticle band structure of monolayer CrBr$_3$ at the $G_0W_0$ level. The indirect quasiparticle band gap is 3.80 eV and the direct band gap at the *M* point is 3.81 eV. Our calculations reveal a strong self-energy correction of 2.13 eV to the direct quasiparticle band gap at the *M* point, due to the weak dielectric screening in reduced dimensions and the localized nature of the Cr *d* states. This self-energy correction is larger than that of 1.77 eV in monolayer CrI$_3$ [28] because the states are more localized with smaller bandwidth in monolayer CrBr$_3$ and the dielectric screening is weaker. Moreover, according to the plot of projected DOS in Fig. 2, we find that self-energy effects push the valence states that are dominated by Br *p* orbitals further down in energy. At the $G_0W_0$ level, majority-spin and minority-spin $t_{2g}$ bands have a similar bandwidth ~0.72 eV. The bandwidth is 0.38 eV for majority-spin $e_g$ bands and 0.34 eV for minority-spin $e_g$ bands. These bandwidths are like those computed at the LSDA+*U* level (see Table S1 of the Supplemental Material [48]). However, self-energy effects on the shape of majority-spin $t_{2g}$ and $e_g$ bands cannot be approximated by a rigid shift, after comparing Figs. 1(e) and 2. The calculated quasiparticle band energies will be used in the next section as input to the BSE calculations.

**IV. Excitonic effects and optical properties**

To solve for exciton eigenstates (correlated electron−hole pairs), we solve the BSE of the interacting two-particle Green's function [31] in the form of an eigenvalue problem:

$$A^S_{cv\mathbf{k}}(\epsilon_{c\mathbf{k}} - \epsilon_{v\mathbf{k}}) + \sum_{c'v'\mathbf{k}'} A^S_{c'v'\mathbf{k}'} \langle cv, \mathbf{k}|\widehat{K}|c'v', \mathbf{k}'\rangle = A^S_{cv\mathbf{k}}\Omega_S, \tag{3}$$

where $\epsilon_{c\mathbf{k}}$ and $\epsilon_{v\mathbf{k}}$ are quasiparticle energies of the conduction and valence bands, respectively. The exciton eigenstate with excitation energy $\Omega_S$ is given by $|S\rangle = \sum_{cv\mathbf{k}} A^S_{cv\mathbf{k}}|cv, \mathbf{k}\rangle$, as a coherent superposition of free electron−hole pairs at different $k$-points. Here, $\widehat{K} = \widehat{K}_d + \widehat{K}_x$ is the electron−hole interaction kernel, containing an attractive direct screened Coulomb term $\widehat{K}_d$ and a repulsive exchange bare Coulomb term $\widehat{K}_x$ [31]. When evaluating $\widehat{K}_d$, we again use Eq. (1) for the irreducible polarizability to keep the BSE matrix Hermitian. We performed the BSE calculation of monolayer CrBr$_3$ within the Tamm-Dancoff approximation [31] and considered interband transitions between 21 valence bands and 14 conduction bands on a $15 \times 15 \times 1$ Monkhorst-Pack $k$-grid to converge the calculation of the optical spectra up to the frequency of 4.2 eV. An energy cutoff of 20 Ry was used for $W$ in constructing the BSE kernel.

The eigenvalues $\Omega_S$ and eigenstates $|S\rangle$ are used to construct the frequency-dependent effective dielectric function tensor $\tilde{\varepsilon}_{\alpha\beta}(\omega)$ in the supercell approach (in units of the vacuum permittivity $\varepsilon_0$):

$$\tilde{\varepsilon}_{\alpha\beta}(\omega) = \delta_{\alpha\beta}\left(1 - \frac{\omega_p^2}{\omega^2}\right) - \frac{1}{\varepsilon_0 \omega^2 N_k V} \sum_S \left[\frac{\langle 0|\hat{j}_p^\alpha|S\rangle\langle S|\hat{j}_p^\beta|0\rangle}{\hbar\omega - \Omega_S + i\eta} - \frac{\langle S|\hat{j}_p^\alpha|0\rangle\langle 0|\hat{j}_p^\beta|S\rangle}{\hbar\omega + \Omega_S + i\eta}\right], \tag{4}$$

where $\omega_p$ is the plasma frequency, $\eta \to 0^+$, and $\alpha, \beta = x, y, z$. Matrix elements of the paramagnetic current operator $\hat{\mathbf{j}}_p = -e\hat{\mathbf{v}}$ between the ground state and a given exciton state $|S\rangle$ are given by

$$\langle 0|\hat{j}_p^\alpha|S\rangle = \sum_{cv\mathbf{k}} A^S_{cv\mathbf{k}} \langle v\mathbf{k}|\hat{j}_p^\alpha|c\mathbf{k}\rangle, \tag{5}$$

where $e$ and $m$ are the elementary charge and electron rest mass, respective, and $\hat{\mathbf{v}} = \frac{1}{i\hbar}[\hat{\mathbf{r}}, \widehat{H}]$ is the single-particle velocity operator.

As an extensive physical quantity, the dielectric function is ill-defined for 2D materials. The meaningful quantity for comparison with physical measurements is the layer polarizability tensor **P**:

$$P_{\alpha\beta} \equiv l(\tilde{\varepsilon}_{\alpha\beta} - \delta_{\alpha\beta})/N_{\text{layer}}, \tag{6}$$

where $l$ is the thickness of the supercell used along the out-of-plane direction and $N_{\text{layer}}$ is the number of layers of the 2D material in a specific calculation. In the following discussion of optical and MO properties, we define the dielectric function of a monolayer CrBr$_3$ using a layer thickness of $d = c_{\text{bulk}}/3 = 6.07$ Å based on its bulk crystal structure [68, 69]:

$$\varepsilon_{\alpha\beta}(\omega) \equiv \delta_{\alpha\beta} + \frac{P_{\alpha\beta}}{d}. \tag{7}$$

It is important to point out that this rescaling process is done to make connection with previous formulations of optical properties of layered systems using dielectric functions. For the optical responses of atomically thin few-layer van der Waals 2D materials, the important physical quantity is **P** which is independent of any assumption of *d*. Since we only consider normal incidence in the following discussion, the in-plane (*xx* and *xy*) components of **P**($\omega$) are calculated and shown in Figs. 3(a) and 3(b). We find that the excitonic effects greatly reshape the dielectric responses by comparing the layer polarizability in Fig. 3(a) (without electron−hole interaction, *GW*-RPA) and those in Fig. 3(b) (with electron−hole interaction, *GW*-BSE). In Fig. 3(c), we calculated the optical absorbance spectrum of linearly polarized light using the diagonal elements of the *GW*-RPA and *GW*-BSE dielectric functions. The *GW*-BSE absorbance spectrum features several absorption peaks below the quasiparticle band gap at ~1.5, 2.1, 2.6 eV, etc. and another strong peak with ~10% absorbance at ~4.0 eV. Several bright exciton states are responsible for these peaks in the absorbance spectrum, as seen in the plot of exciton energy levels in Fig. 4(a), where the bright exciton states are colored in red and dark states in gray. Unlike monolayer CrI$_3$, the top valence bands are flat in monolayer CrBr$_3$, creating a large joint DOS across the band gap which strongly enhances the excitonic effects and thus increases the exciton binding energies. The first bright exciton has an excitation energy of 1.5 eV, which leads to a huge binding energy of 2.3 eV. In the following, we visualize the exciton amplitudes in real space in Figs. 4(b)−4(e). Note that the valence and conduction Bloch waves here are all two-component spinor wavefunctions, which means the electron−hole distribution in real space for a selected exciton state should be calculated as

$$A_S(\mathbf{r}_e, \mathbf{r}_h) \equiv \sum_{cv\mathbf{k}} |A_{cv\mathbf{k}}^S|^2 \sum_{\sigma_h} |\phi_{v\mathbf{k}}(\mathbf{r}_h, \sigma_h)|^2 \sum_{\sigma_e} |\phi_{c\mathbf{k}}(\mathbf{r}_e, \sigma_e)|^2, \tag{8}$$

where $\sigma_h$ and $\sigma_e$ refers to the *z*-axis spinor components of the hole ($\phi_{v\mathbf{k}}^*$) and electron ($\phi_{c\mathbf{k}}$) band states, respectively. By calculating the expectation value of spin operators for the electron and hole in an exciton state, we find that the bright exciton states in Figs. 4(b)−4(e) all consist of majority-spin electrons and minority-spin holes (i.e., formed by up-to-up interband transitions, see Sec. II of the Supplemental Material [48]). Unlike monolayer transition metal dichalcogenides where the lowest-energy bright excitons are of the Wannier–Mott type with a diameter of several nanometers [34, 35], ferromagnetic monolayer CrBr$_3$ hosts bright charge-transfer exciton states that extends over one to several primitive cells ($\leq$ 1 nm) which is still large compared with atomic size and is indicative of excitonic formation from band transitions instead of intra-atomic $d$–$d$ transitions. These plots of exciton distribution are also consistent with the intuition that a larger exciton binding energy is related to a smaller exciton radius [70].

**V. MO effects**

In magneto-optics, a linearly polarized continuous-wave light propagating through a medium is modified by the presence of a magnetic field, where the $\sigma^+$ and $\sigma^-$ circularly polarized components propagate with different refractive index and therefore pick up different optical path length and absorption. There are several important MO effects, such as the Faraday effect [71], the MO Kerr effect [72], and MCD. In the Kerr effect, the polarization change of the reflected light is measured, while in the Faraday effect, the polarization change of the transmitted light is measured. That is, the reflected or transmitted light become elliptically polarized (characterized by an ellipticity angle) and the long axis of the polarization ellipse is rotated (characterized by a rotation angle). In this paper, we consider the most common MO setup used to study 2D magnets [5, 9]: the polar setup with normal incidence, where the direction of magnetization is parallel to the out-of-plane direction ($+z$). MCD, on the other hand, describes differential absorption of circularly polarized lights. SOC effects are important to achieve non-zero MO effects in rhombohedral systems such as CrBr$_3$, because they break the orbital degeneracy such that optical transitions corresponding to different circularly polarized lights are no longer equivalent [24–26, 73].

For a ferromagnetic material with $C_3$ rotational symmetry along the spin-polarization direction ($z$-axis), its frequency-dependent layer polarizability tensor as a function of the magnetic field takes the following form:

$$\mathbf{P}(\omega, \mathbf{B}) = \begin{pmatrix} P_{xx}(\omega, \mathbf{B}) & P_{xy}(\omega, \mathbf{B}) & 0 \\ -P_{xy}(\omega, \mathbf{B}) & P_{xx}(\omega, \mathbf{B}) & 0 \\ 0 & 0 & P_{zz}(\omega, \mathbf{B}) \end{pmatrix}, \qquad (9)$$

where **B** is the internal Weiss field and we have applied $P_{yy} = P_{xx}, P_{yx} = -P_{xy}$, and $P_{xz} = P_{zx} = P_{yz} = P_{zy} = 0$ due to the $C_3$ symmetry (see the Appendix). To simplify the expression of the diagonal element $P_{xx}$, we calculate the imaginary part of $P_{xx}$ in the limit of $\eta \to 0^+$,

$$\text{Im } P_{xx}(\omega) = \frac{\pi \hbar^2 l}{\varepsilon_0 N_k V N_{\text{layer}}} \sum_S \frac{1}{\Omega_S^2} |\langle 0|\hat{j}_p^x|S\rangle|^2 [\delta(\hbar\omega - \Omega_S) - \delta(\hbar\omega + \Omega_S)], \quad (10)$$

where we have replaced $1/\omega^2$ by $\hbar^2/\Omega_S^2$ due to the Dirac delta functions. In the limit of low frequency, the Thomas–Reiche–Kuhn sum rule $\omega_p^2 = \frac{2}{\varepsilon_0 N_k V} \sum_S \frac{|\langle 0|\hat{j}_p^\alpha|S\rangle|^2}{\Omega_S}$ leads to the cancellation of the frequency-dependent parts of both the first and second terms in Eq. (4), leading to a finite value of the dielectric constant at zero frequency, as expected for a semiconductor. Then the expression of the real part of $P_{xx}$ can be derived with the Kramers−Kronig relation:

$$\text{Re } P_{xx}(\omega) = \frac{1}{\pi} \mathcal{P} \int_{-\infty}^{\infty} d\omega' \frac{\text{Im } P_{xx}(\omega')}{\omega' - \omega}$$

$$= -\frac{\hbar^2 l}{\varepsilon_0 N_k V N_{\text{layer}}} \sum_S \frac{1}{\Omega_S^2} |\langle 0|\hat{j}_p^x|S\rangle|^2 \left[\frac{1}{\hbar\omega - \Omega_S} - \frac{1}{\hbar\omega + \Omega_S}\right], \quad (11)$$

which is now numerically stable around $\omega = 0$. Combining Eqs. (4)–(6), and (9), we can prove that $\text{Re } \langle 0|\hat{j}_p^x|S\rangle\langle S|\hat{j}_p^y|0\rangle \equiv 0$ (see the Appendix), which means the real part of $P_{xy}$ is given by

$$\text{Re } P_{xy}(\omega) = \frac{i\pi \hbar^2 l}{\varepsilon_0 N_k V N_{\text{layer}}} \sum_S \frac{1}{\Omega_S^2} [\langle 0|\hat{j}_p^x|S\rangle\langle S|\hat{j}_p^y|0\rangle \delta(\hbar\omega - \Omega_S)$$

$$- \langle S|\hat{j}_p^x|0\rangle\langle 0|\hat{j}_p^y|S\rangle \delta(\hbar\omega + \Omega_S)]. \quad (12)$$

Then the expression of the imaginary part of $P_{xy}$ can be derived with the Kramers−Kronig relation:

$$\text{Im } P_{xy}(\omega) = -\frac{1}{\pi} \mathcal{P} \int_{-\infty}^{\infty} d\omega' \frac{\text{Re } P_{xy}(\omega')}{\omega' - \omega}$$

$$= \frac{i\hbar^2 l}{\varepsilon_0 N_k V N_{\text{layer}}} \sum_S \frac{1}{\Omega_S^2} \left[\frac{\langle 0|\hat{j}_p^x|S\rangle\langle S|\hat{j}_p^y|0\rangle}{\hbar\omega - \Omega_S} - \frac{\langle S|\hat{j}_p^x|0\rangle\langle 0|\hat{j}_p^y|S\rangle}{\hbar\omega + \Omega_S}\right]. \quad (13)$$

To model the inhomogeneous broadening observed in optical experiments, we add a Gaussian broadening, $\delta(\hbar\omega \pm \Omega_S) \approx \frac{1}{\eta\sqrt{2\pi}} e^{-\frac{(\hbar\omega \pm \Omega_S)^2}{2\eta^2}}$ to Eqs. (10) and (12). A regularization procedure $\frac{1}{\hbar\omega \pm \Omega_S} \approx \frac{\hbar\omega \pm \Omega_S}{(\hbar\omega \pm \Omega_S)^2 + \eta^2}$ is used to avoid numerical problems with fine frequency sampling for Eqs. (11) and (13). The validity of Eqs. (10)–(13) has been verified by previous work [55].

We define the complex refractive index with the direction of the wave vector **k** of light, $\mathbf{n} \equiv \frac{c\mathbf{k}}{\omega}$, where $c$ is the speed of light in vacuum and $\omega$ is the frequency of light. By solving the wave equation $(n^2 \mathbb{I} - \boldsymbol{\varepsilon} - \mathbf{n}\mathbf{n}^\top) \cdot \mathbf{E} = 0$ with Eqs. (7) and (9), we get the eigenmodes as $\sigma^+$ and $\sigma^-$ circularly polarized plane waves, with distinct refractive indices:

$$(n_\pm(\omega, \mathbf{B}))^2 = \varepsilon_{xx}(\omega, \mathbf{B}) \pm i\varepsilon_{xy}(\omega, \mathbf{B}), \tag{14}$$

of which the complex electric field amplitude points along the direction of the basis, $\hat{\mathbf{e}}_\pm = \frac{\mp}{\sqrt{2}}(\hat{\mathbf{e}}_x \pm i\hat{\mathbf{e}}_y)$. In the wave equation, $\mathbf{n}$ is a column vector and $\mathbf{n}^\top$ is the transposed row vector. In this paper, we avoid the ambiguous terminology of left and right circularly polarized lights and stick to the usage of $\sigma^+$ and $\sigma^-$ circularly polarized lights, which are well-defined after the Cartesian coordinate system is set up. Moreover, according to the Onsager reciprocal relations [74], we have

$$\varepsilon_{\alpha\beta}(\omega, \mathbf{B}) = \varepsilon_{\beta\alpha}(\omega, -\mathbf{B}). \tag{15}$$

Equation (15) allows us to easily calculate the MO effects when the magnetization direction is flipped. If there is no net magnetization or external magnetic field (i.e., $\mathbf{B} = 0$), Eqs. (7), (9), and (15) lead to $\varepsilon_{xy} \equiv 0$, which means there are no MO effects.

Using Eq. (14), we calculated the absorbance spectrum of circularly polarized lights in Fig. 5(a) without electron−hole interaction, and in Fig. 5(b) with electron−hole interaction. Note that the ranges of frequency and absorbance are different in these two plots. Comparing Figs. 5(a) and 5(b), we find that the electron−hole interaction enhances the optical absorption for both circularly polarized lights in the frequency range of interest. In addition, by calculating the MCD signal as $(A_+ - A_-)/(A_+ + A_-)$ in Fig. 5(c), where $A_\pm$ denotes the absorbance for $\sigma^\pm$ circularly polarized lights, we find that the excitonic effects significantly enhance the dichroism over a large frequency range up to the onset of the electron−hole continuum. This is because the BSE matrix, which involves the dielectric function and carrier wavefunctions in the presence of a magnetic order, has broken time-reversal symmetry. Therefore, the excitonic effects are different on different circularly polarized transitions. The peak position and helicity of the first bright exciton peak in Fig. 5(b) agree well with recent polarization-resolved magneto-photoluminescence measurements, where the photoluminescence peak is located at 1.35 eV [41].

In the following, we discuss how to calculate MO Kerr and Faraday signals in a two-interface polar setup, as shown in Fig. 6(a). The system consists of two interfaces located at $z_0$ and $z_1$, with the middle (first) layer being the atomically thin magnetic material of interest. The left (zeroth) and right (second) layers are semi-infinitely thick. A normal incident light comes from the right side, and we measure the reflected light on the right side or the transmitted light on the left side. To mathematically describe how an electromagnetic wave interacts with such stratified and anisotropic media, we adopt a 4×4 formalism involving the in-plane components of both electric ($E_x, E_y$) and magnetic fields ($B_x, B_y$). This formalism has been used to study birefringent multilayer media and MO ellipsometry [75, 76], and it can be generalized to more complex setups. We consider a ferromagnetic monolayer CrBr$_3$ magnetized along the +z direction, and the second layer is vacuum. Within each layer ($l = 0, 1, 2$), we choose the four eigenmodes of light as follows: $\hat{\mathbf{e}}^{(l)}_1 = \hat{\mathbf{e}}^{(l)}_2 = \hat{\mathbf{e}}_+, \hat{\mathbf{e}}^{(l)}_3 = \hat{\mathbf{e}}^{(l)}_4 = \hat{\mathbf{e}}_-$, with the corresponding refractive indices: $n^{(l)}_1 = -n^{(l)}_2 = n_+$, $n^{(l)}_3 = -n^{(l)}_4 = n_-$. The electric and magnetic fields of light in the first and second layers are given by

$$\mathbf{E}^{(l)} = \sum_{j=1}^{4} E^{(l)}_{0j} \hat{\mathbf{e}}^{(l)}_j e^{i\left(k^{(l)}_j (z-z_{l-1}) - \omega t\right)}, \tag{16}$$

and

$$c\mathbf{B}^{(l)} = \sum_{j=1}^{4} E^{(l)}_{0j} \mathbf{b}^{(l)}_j e^{i\left(k^{(l)}_j (z-z_{l-1}) - \omega t\right)}, \tag{17}$$

with $\mathbf{b}^{(l)}_j = n^{(l)}_j \hat{\mathbf{e}}_z \times \hat{\mathbf{e}}^{(l)}_j$, and $k^{(l)}_j = \frac{\omega}{c} n^{(l)}$. The electric and magnetic fields of light in the zeroth layer are given by

$$\mathbf{E}^{(0)} = \sum_{j=1}^{4} E^{(0)}_{0j} \hat{\mathbf{e}}^{(0)}_j e^{i\left(k^{(0)}_j (z-z_0) - \omega t\right)}, \tag{18}$$

and

$$c\mathbf{B}^{(0)} = \sum_{j=1}^{4} E^{(0)}_{0j} \mathbf{b}^{(0)}_j e^{i\left(k^{(0)}_j (z-z_0) - \omega t\right)}. \tag{19}$$

The requirement of the continuity of the tangential field components at the interfaces connects the field amplitudes $E^{(l)}_{0j}$ between two layers. The dynamical matrix within each layer is given by a block-diagonal form:

$$\mathbf{D}^{(l)} = \begin{bmatrix} \hat{\mathbf{e}}_1^{(l)} \cdot \hat{\mathbf{e}}_+^* & \hat{\mathbf{e}}_2^{(l)} \cdot \hat{\mathbf{e}}_+^* & \hat{\mathbf{e}}_3^{(l)} \cdot \hat{\mathbf{e}}_+^* & \hat{\mathbf{e}}_4^{(l)} \cdot \hat{\mathbf{e}}_+^* \\ \mathbf{b}_1^{(l)} \cdot \hat{\mathbf{e}}_+^* & \mathbf{b}_2^{(l)} \cdot \hat{\mathbf{e}}_+^* & \mathbf{b}_3^{(l)} \cdot \hat{\mathbf{e}}_+^* & \mathbf{b}_4^{(l)} \cdot \hat{\mathbf{e}}_+^* \\ \hat{\mathbf{e}}_1^{(l)} \cdot \hat{\mathbf{e}}_-^* & \hat{\mathbf{e}}_2^{(l)} \cdot \hat{\mathbf{e}}_-^* & \hat{\mathbf{e}}_3^{(l)} \cdot \hat{\mathbf{e}}_-^* & \hat{\mathbf{e}}_4^{(l)} \cdot \hat{\mathbf{e}}_-^* \\ \mathbf{b}_1^{(l)} \cdot \hat{\mathbf{e}}_-^* & \mathbf{b}_2^{(l)} \cdot \hat{\mathbf{e}}_-^* & \mathbf{b}_3^{(l)} \cdot \hat{\mathbf{e}}_-^* & \mathbf{b}_4^{(l)} \cdot \hat{\mathbf{e}}_-^* \end{bmatrix} = \begin{bmatrix} 1 & 1 & 0 & 0 \\ -i\,n_+^{(l)} & i\,n_+^{(l)} & 0 & 0 \\ 0 & 0 & 1 & 1 \\ 0 & 0 & i\,n_-^{(l)} & -i\,n_-^{(l)} \end{bmatrix}. \quad (20)$$

The propagation matrix is defined as a diagonal matrix with entries being the phase shift associated with the optical path length for each eigenmode within the sample, $\mathbf{P}^{(1)} = \mathrm{diag}\{e^{i\delta_+}, e^{-i\delta_+}, e^{i\delta_-}, e^{-i\delta_-}\}$, where $\delta_\pm = \frac{\omega}{c} n_\pm d$ and $d$ is the thickness of the middle layer. In this two-interface setup, $\mathbf{E}_0^{(0)}$ and $\mathbf{E}_0^{(2)}$ are related by the transfer matrix $\mathbf{M}_c$ in the basis of circularly polarized lights:

$$\mathbf{E}_0^{(2)} = \mathbf{M}_c \mathbf{E}_0^{(0)} = \left[ \left(\mathbf{D}^{(2)}\right)^{-1} \mathbf{D}^{(1)} \mathbf{P}^{(1)} \left(\mathbf{D}^{(1)}\right)^{-1} \mathbf{D}^{(0)} \right] \mathbf{E}_0^{(0)}. \quad (21)$$

Here, $\mathbf{M}_c$ has a simple block-diagonal form:

$$\begin{aligned} \mathbf{M}_c &= \begin{bmatrix} \mathbf{M}_+ & 0 \\ 0 & \mathbf{M}_- \end{bmatrix}, \\ \mathbf{M}_\pm &= \frac{1}{t_{21}^\pm t_{10}^\pm} \begin{bmatrix} e^{i\delta_\pm} + e^{-i\delta_\pm} r_{21}^\pm r_{10}^\pm & e^{i\delta_\pm} r_{10}^\pm + e^{-i\delta_\pm} r_{21}^\pm \\ e^{i\delta_\pm} r_{21}^\pm + e^{-i\delta_\pm} r_{10}^\pm & e^{i\delta_\pm} r_{21}^\pm r_{10}^\pm + e^{-i\delta_\pm} \end{bmatrix}, \end{aligned} \quad (22)$$

where $r_{mn}$ and $t_{mn}$ are the one-interface Fresnel coefficients from the $m$-th layer to the $n$-th layer. However, in the measurement of MO signals, we use a linearly polarized light and measure the polarization of the reflected or transmitted elliptically polarized light. Assuming the left and right layers are non-magnetic and isotropic with one optic axis of their dielectric function tensor pointing along the $z$ direction, we adopt a basis transformation from the circularly polarized light to linearly polarized light in the left and right layers: $\hat{\mathbf{e}}_1^{(l)} = \hat{\mathbf{e}}_2^{(l)} = \hat{\mathbf{e}}_x$, $\hat{\mathbf{e}}_3^{(l)} = \hat{\mathbf{e}}_4^{(l)} = \hat{\mathbf{e}}_y$, as well as $n_1^{(l)} = -n_2^{(l)} = n_3^{(l)} = -n_4^{(l)} = n^{(l)}$, for $l = 0, 2$. In the following, this new basis of linearly polarized plane waves is denoted as $\{x \rightarrow, x \leftarrow, y \rightarrow, y \leftarrow\}$, emphasizing the polarization and propagation direction of each mode. In this basis of linearly polarized lights, the electric field amplitudes in the left and right layer are related by transfer matrix $\mathbf{M}$:

$$\begin{bmatrix} E_{0x\rightarrow}^{(2)} \\ E_{0x\leftarrow}^{(2)} \\ E_{0y\rightarrow}^{(2)} \\ E_{0y\leftarrow}^{(2)} \end{bmatrix} = \mathbf{M} \begin{bmatrix} E_{0x\rightarrow}^{(0)} \\ E_{0x\leftarrow}^{(0)} \\ E_{0y\rightarrow}^{(0)} \\ E_{0y\leftarrow}^{(0)} \end{bmatrix}, \quad (23)$$

$$\mathbf{M} = \frac{1}{2}\begin{bmatrix} \mathbf{M}_+ + \mathbf{M}_- & -i(\mathbf{M}_+ - \mathbf{M}_-) \\ i(\mathbf{M}_+ - \mathbf{M}_-) & \mathbf{M}_+ + \mathbf{M}_- \end{bmatrix}.$$

As mentioned above, we consider an incoming $x$-polarized light from the second medium to the zeroth medium, that is, $E_{0y\leftarrow}^{(2)} = 0$. Moreover, there are no incident lights from the zeroth medium to the second medium, which means, $E_{0x\rightarrow}^{(0)} = E_{0y\rightarrow}^{(0)} = 0$. With these conditions, we can calculate the reflectivities and transmissivities as follows:

$$t_{ss} \equiv \frac{E_{0x\leftarrow}^{(0)}}{E_{0x\leftarrow}^{(2)}} = \frac{M_{44}}{M_{22}M_{44} - M_{24}M_{42}}, \tag{24}$$

$$t_{sp} \equiv \frac{E_{0y\leftarrow}^{(0)}}{E_{0x\leftarrow}^{(2)}} = \frac{-M_{42}}{M_{22}M_{44} - M_{24}M_{42}}, \tag{25}$$

$$r_{ss} \equiv \frac{E_{0x\rightarrow}^{(2)}}{E_{0x\leftarrow}^{(2)}} = \frac{M_{12}M_{44} - M_{14}M_{42}}{M_{22}M_{44} - M_{24}M_{42}}, \tag{26}$$

$$r_{sp} \equiv \frac{E_{0y\rightarrow}^{(2)}}{E_{0x\leftarrow}^{(2)}} = \frac{M_{32}M_{44} - M_{34}M_{42}}{M_{22}M_{44} - M_{24}M_{42}}. \tag{27}$$

As shown in Fig. 6(b), the Faraday signals for the transmitted (left-moving) electric field $\mathbf{E}_\leftarrow^{(0)} = E_{0x\leftarrow}^{(0)} \hat{\mathbf{e}}_x + E_{0y\leftarrow}^{(0)} \hat{\mathbf{e}}_y$, are determined by the ratio of transmissitivities $t_{sp}/t_{ss}$ [77]:

$$\tan 2\theta_\mathrm{F} = \frac{2\left|\frac{t_{sp}}{t_{ss}}\right| \cos\left(\arg \frac{t_{sp}}{t_{ss}}\right)}{1 - \left|\frac{t_{sp}}{t_{ss}}\right|^2}, \quad -\frac{\pi}{2} < \theta_\mathrm{F} \leq \frac{\pi}{2}, \tag{28}$$

$$\sin 2\chi_\mathrm{F} = \frac{2\left|\frac{t_{sp}}{t_{ss}}\right| \sin\left(\arg \frac{t_{sp}}{t_{ss}}\right)}{1 + \left|\frac{t_{sp}}{t_{ss}}\right|^2}, \quad -\frac{\pi}{4} < \chi_\mathrm{F} \leq \frac{\pi}{4}. \tag{29}$$

Similarly, the Kerr signals for the reflected (right-moving) electric field $\mathbf{E}_\rightarrow^{(2)} = E_{0x\rightarrow}^{(2)} \hat{\mathbf{e}}_x + E_{0y\rightarrow}^{(2)} \hat{\mathbf{e}}_y$, are determined by the ratio of reflectivities $r_{sp}/r_{ss}$:

$$\tan 2\theta_\mathrm{K} = \frac{2\left|\frac{r_{sp}}{r_{ss}}\right| \cos\left(\arg \frac{r_{sp}}{r_{ss}}\right)}{1 - \left|\frac{r_{sp}}{r_{ss}}\right|^2}, \quad -\frac{\pi}{2} < \theta_\mathrm{K} \leq \frac{\pi}{2}, \tag{30}$$

$$\sin 2\chi_K = \frac{2\left|\frac{r_{sp}}{r_{ss}}\right|\sin(\arg\frac{r_{sp}}{r_{ss}})}{1+\left|\frac{r_{sp}}{r_{ss}}\right|^2}, -\frac{\pi}{4} < \chi_K \le \frac{\pi}{4}. \tag{31}$$

Next, we apply Eqs. (24)–(31) to study the MO Kerr and Faraday signals of monolayer CrBr$_3$ with different substrate materials. First, we consider the simplest case with both left and right layers being just vacuum and calculate the Faraday angle $\theta_F$ (solid blue) and ellipticity $\chi_F$ (dashed red) in Fig. 6(d). We find that $\theta_F$ and $\chi_F$ are connected through a set of approximate Kramers−Kronig relations, as expected from previous work [24]. As shown in Figs. 6(d)−6(f), we find a strong dependence of the MO signals on the excitation frequency. We predict a maximal positive $\theta_F$ of 2 mrad around the excitation frequency of 2.6 eV and negative $\theta_F$ of the order of −1 mrad between 2.0 and 2.5 eV. Furthermore, we calculate the Kerr signals in the presence of conventional thick substrate materials, namely, sapphire and fused silica. Both insulating substrates have large band gaps and little absorption in the range of 1.0−3.5 eV. In this paper, we modeled their refractive indices using experimental values at the relevant frequencies with $n = 1.5$ for fused silica [78] and $n = 1.75$ for sapphire [79]. In Figs. 6(e) and 6(f), we show the calculated Kerr signals in the setup with sapphire and fused silica substrates, respectively. We find a similar shape for the signals but with opposite sign compared with the Faraday signals in Fig. 6(d). In addition, different substrate materials can strongly modify the amplitudes of Kerr signals. For example, the signals in the fused silica setup are almost twice as large those in the sapphire setup. Close attention therefore should be paid in interpreting MO experiments on atomically thin 2D magnetic semiconductors with different substrate configurations.

## VI. Summary

In conclusion, we present a detailed theoretical formalism to model the optical and MO properties of 2D magnetic semiconductors, including the important SOC and excitonic effects. The theoretical and numerical methods presented in this paper can also be applied to other mono- or multi-layer 2D magnets, as well as van der Waals heterostructures consisting of 2D magnets. Using the first-principles full-spinor *GW* and *GW*-BSE methods without time-reversal symmetry, we calculate the exciton eigenstates, layer polarizability tensor, as well as optical and MO spectra of a prototypical 2D magnetic semiconductor, ferromagnetic monolayer CrBr$_3$. The calculated optical absorbance spectra and MO signals demonstrate dominant excitonic effects. With a two-interface model, we also find that the substrate refractive index will significantly affect the MO signals. In this paper, we provide a theoretical framework and a first-principles approach to simulate the optical and MO properties of 2D magnetic semiconductors and shed light on possible design principles for building optoelectronic and spintronic devices with magnetic van der Waals materials.

**Acknowledgements**

The paper was supported by the Theory Program at the Lawrence Berkeley National Lab (LBNL) through the Office of Basic Energy Sciences, U.S. Department of Energy (DOE) under Contract No. DE-AC02-05CH11231, which provided the *GW* and *GW*-BSE calculations and simulations as well as the theoretical formulation and analysis of the MO effects. Advanced codes were provided by the Center for Computational Study of Excited-State Phenomena in Energy Materials at LBNL, which is funded by the U.S. DOE, Office of Science, Basic Energy Sciences, Materials Sciences and Engineering Division under Contract No. DE-AC02-05CH11231, as part of the Computational Materials Sciences Program. Computational resources were provided by the DOE at LBNL's NERSC facility and by the National Science Foundation (Grant No. DMR-1926004) through XSEDE Bridges-2 system at the Pittsburgh Supercomputing Center and through TACC Frontera system at the University of Texas at Austin.

**Appendix**

In this section, we prove the form of the matrix in Eq. (9). Monolayer CrBr$_3$ has $S_6 = C_3 \otimes C_i$ point group symmetry, and a general layer polarizability tensor is given by

$$\mathbf{P} = \begin{pmatrix} P_{xx} & P_{xy} & P_{xz} \\ P_{yx} & P_{yy} & P_{yz} \\ P_{zx} & P_{zy} & P_{zz} \end{pmatrix}. \tag{A1}$$

Here, **P** should be invariant under the $C_3$ rotational operation given by a SO(3) rotation matrix **D**:

$$\mathbf{D} = \begin{pmatrix} -\frac{1}{2} & -\frac{\sqrt{3}}{2} & 0 \\ \frac{\sqrt{3}}{2} & -\frac{1}{2} & 0 \\ 0 & 0 & 1 \end{pmatrix}, \tag{A2}$$

which leads to,

$$\mathbf{P} \equiv \mathbf{D}\mathbf{P}\mathbf{D}^{-1}. \tag{A3}$$

Comparing each entry of the matrices on both sides of Eq. (A3), we have the following identities:

$$P_{xz} = P_{zx} = P_{yz} = P_{zy} = 0, \tag{A4}$$

$$P_{xx} = P_{yy}, \tag{A5}$$

$$P_{yx} = -P_{xy}. \tag{A6}$$

Combining Eqs. (4)–(6) and (A6), we arrive at

$$\sum_S [\langle 0|\hat{j}_p^x|S\rangle\langle S|\hat{j}_p^y|0\rangle + \langle 0|\hat{j}_p^y|S\rangle\langle S|\hat{j}_p^x|0\rangle] \left[\frac{1}{\hbar\omega - \Omega_S + i\eta} - \frac{1}{\hbar\omega + \Omega_S + i\eta}\right] \equiv 0, \forall \omega. \tag{A7}$$

Equation (A7) leads to

$$\langle 0|\hat{j}_p^x|S\rangle\langle S|\hat{j}_p^y|0\rangle + \langle 0|\hat{j}_p^y|S\rangle\langle S|\hat{j}_p^x|0\rangle = 2\mathrm{Re}\,\langle 0|\hat{j}_p^x|S\rangle\langle S|\hat{j}_p^y|0\rangle \equiv 0, \forall S. \tag{A8}$$

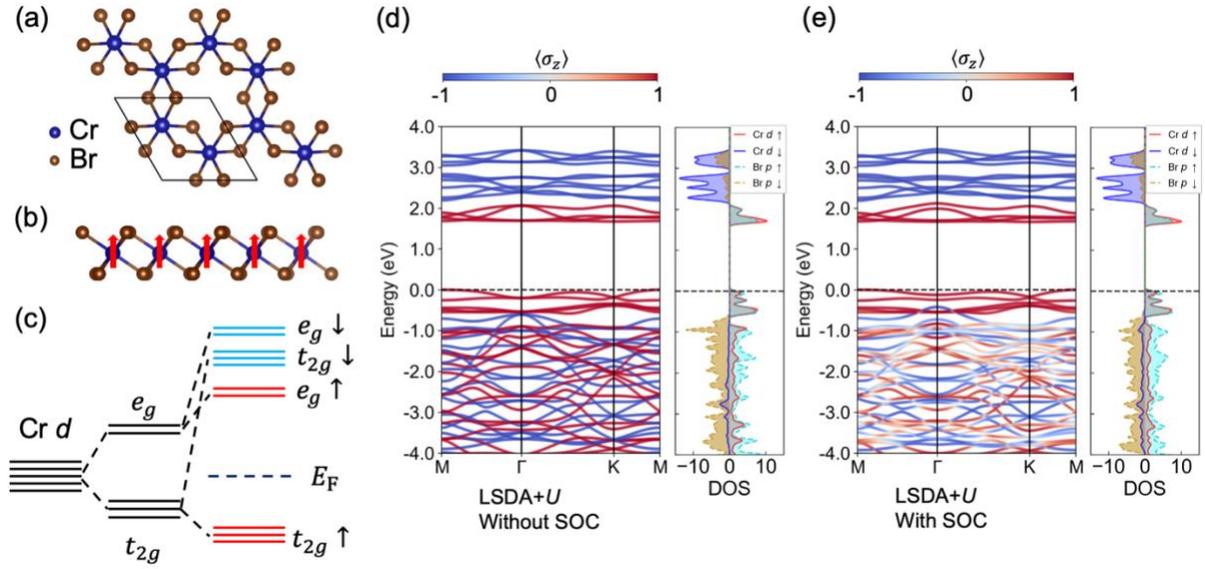

FIG. 1. Crystal structure and electronic structure of ferromagnetic monolayer CrBr$_3$. (a) Top view and (b) side view of the crystal structure of monolayer CrBr$_3$. Cr atoms are in blue while Br atoms in brown. Red arrows denote the out-of-plane magnetization, which points along the $+z$ direction. (c) Schematic energy diagrams of Cr $d$ orbitals in the presence of an octahedral crystal field and magnetic exchange interaction. The horizontal dashed line denotes the Fermi level. Local-spin-density approximation with an onsite Hubbard potential (LSDA + $U$) band structure (left) and projected DOS (right) of monolayer CrBr$_3$ (d) without and (e) with spin-orbit coupling (SOC) effects. A rotationally invariant Hubbard potential is employed with $U = 1.5$ eV and $J = 0.5$ eV in the LSDA + $U$ calculation. Colors denote the magnitude of spin polarization along the out-of-plane direction. The red (blue) color denotes the majority-spin (minority-spin) polarization. The DOS (in units of states per electronvolt per unit cell) is decomposed into contributions from Cr majority-spin 3$d$ orbitals (red curve), Cr minority-spin 3$d$ orbitals (blue curve), Br majority-spin 4$p$ orbitals (cyan curve) and Br minority-spin 4$p$ orbitals (brown curve). The energy of the valence band maximum is set to zero. A Gaussian broadening of 50 meV is used for the projected DOS plots.

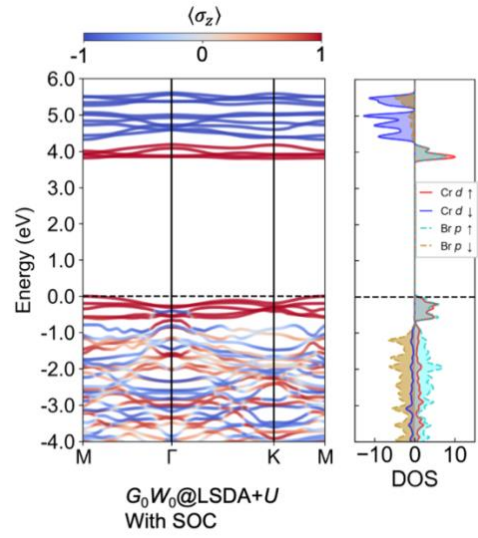

FIG. 2. $G_0W_0$ band structure (left) and projected density of states (DOS) (right) of monolayer CrBr$_3$ with spin-orbit coupling (SOC) effects. Computational parameters and the color scheme are the same as in Fig. 1.

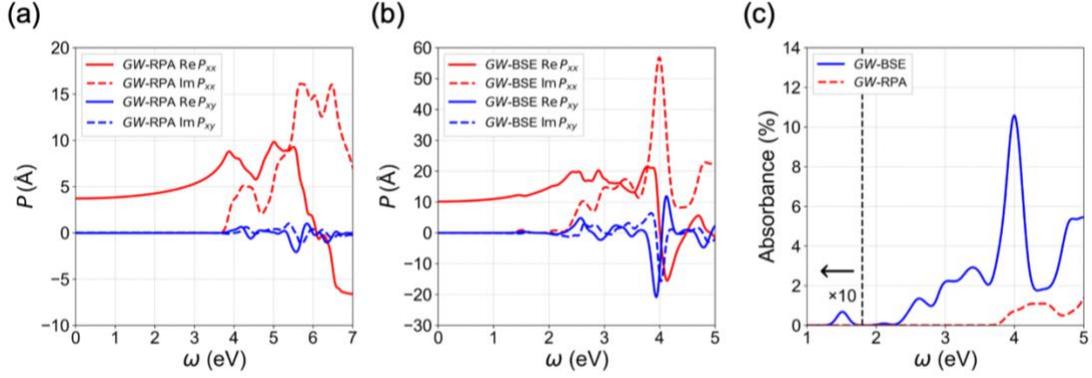

FIG. 3. (a) Calculated real part (solid lines) and imaginary part (dashed lines) of the diagonal $P_{xx}$ (red) and off-diagonal $P_{xy}$ (blue) layer polarizability of ferromagnetic monolayer CrBr$_3$ without electron−hole interaction ($GW$-RPA). (b) Calculated real part (solid lines) and imaginary part (dashed lines) of the diagonal $P_{xx}$ (red) and off-diagonal $P_{xy}$ (blue) with electron−hole interaction ($GW$-BSE). (c) Absorbance spectrum of linearly polarized light with electron−hole interaction ($GW$-BSE, solid blue line) and without electron−hole interaction ($GW$-RPA, dashed red line). The amplitudes <1.8 eV (indicated by black dashed line) are multiplied by 10 for better visibility. An 80 meV energy broadening is applied.

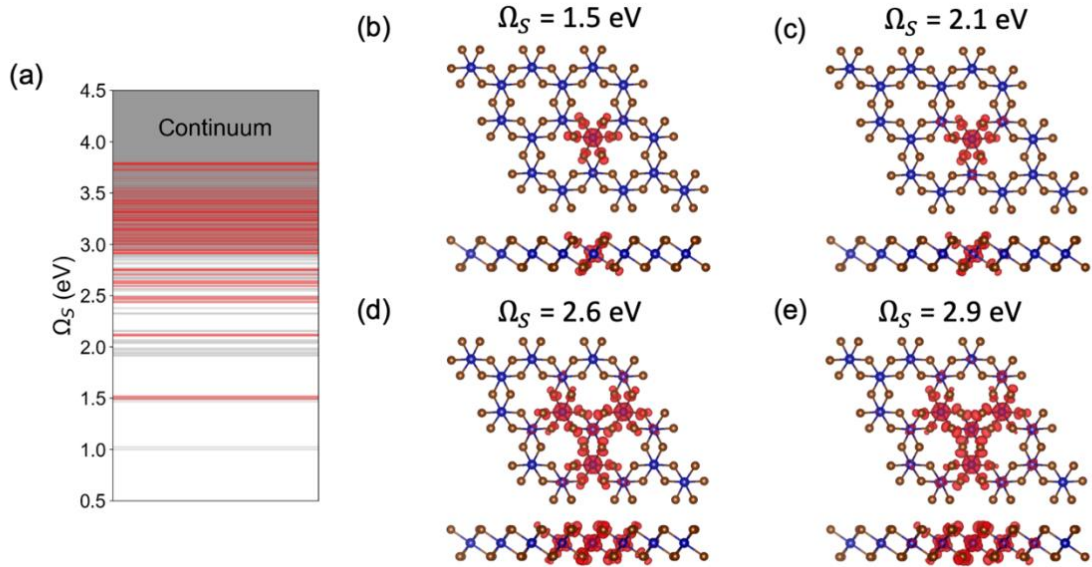

FIG. 4. (a) Exciton energy levels of ferromagnetic monolayer CrBr$_3$ calculated using the first-principles *GW*-BSE method. Energy levels of optically bright exciton states are in red, while those of dark states are in gray. The bright excitons have at least two orders of magnitude stronger oscillator strength compared with the dark ones. The free electron−hole continuum starts from 3.81 eV. (b)-(e) Top view and side view of electron distribution in real space with the hole fixed on a Cr atom for selected bright exciton states. The exciton excitation energies are: (b) 1.5 eV, (c) 2.1 eV, (d) 2.6 eV, and (e) 2.9 eV. Shown are isovalue surfaces with the value set at 1% of the maximum. Here, the dominant states (with the largest oscillator strength among the nearby bright states) are plotted.

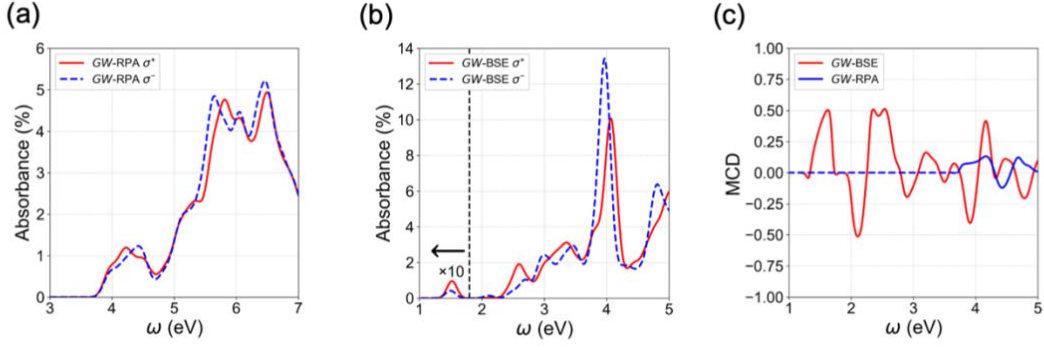

FIG. 5. (a) Absorbance spectrum of circularly polarized lights without electron−hole interaction (*GW*-RPA). (b) Absorbance spectrum of circularly polarized lights with electron−hole interaction (*GW*-BSE). The solid red (dashed blue) curve corresponds to the $\sigma^+$ ($\sigma^-$) circularly polarized light. The amplitudes <1.8 eV (black dashed line) are multiplied by 10 for better visibility in (b). (c) Frequency-dependent magnetic circular dichroism (MCD) signal of absorbance with (red, *GW*-BSE) and without (blue, *GW*-RPA) electron−hole interaction. The MCD signal is calculated as $(A_+ - A_-)/(A_+ + A_-)$, where $A_\pm$ denotes the absorbance for $\sigma^\pm$ circularly polarized lights in (a) and (b). To avoid numerical instability, a small imaginary part of $\eta = 0.001i$ is added to the denominator and the real part is plotted.

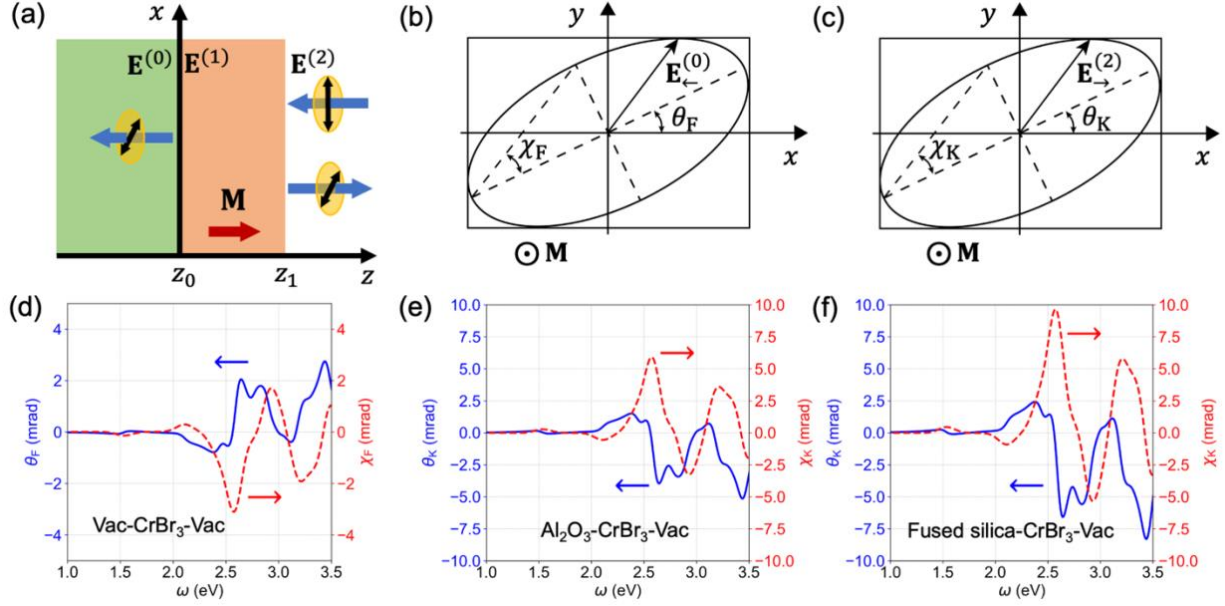

FIG. 6. (a) Polar setup of the magneto-optical (MO) effects with two interfaces (located at $z_0$ and $z_1$). Each layer is homogeneous in the *x-y* plane. The left (zeroth) and right (second) media are semi-infinitely thick, and the middle (first) layer has a finite thickness $d = z_1 - z_0$. The normal incident light coming from right to left is linearly polarized along the *x*-axis. The red arrow pointing along the +*z* direction denotes the magnetization of layer 1. $\mathbf{E}^{(0)}$ and $\mathbf{E}^{(1)}$ denote the amplitudes of electric fields in the zeroth and first layer at the $z_0$ interface, respectively. $\mathbf{E}^{(2)}$ denotes the amplitude of electric field in the second layer at the $z_1$ interface. (b) The polarization plane of the transmitted light. $\mathbf{E}^{(0)}_{\leftarrow}$ denotes the electric field amplitude of the transmitted (left-moving) light. The polarization ellipse is oriented at a Faraday angle $\theta_F$ with respect to the *x*-axis. The Faraday ellipticity is defined through the Faraday ellipticity angle $\chi_F$. (b) The polarization plane of the reflected light. $\mathbf{E}^{(2)}_{\rightarrow}$ denotes the electric field amplitude of the reflected (right-moving) light. The polarization ellipse is oriented at a Kerr angle $\theta_K$ with respect to the *x*-axis. The Kerr ellipticity is defined through the ellipticity angle $\chi_K$. (d) Calculated Faraday angle and ellipticity in the setup with layers of semi-infinite vacuum, monolayer CrBr3, and semi-infinite vacuum. (e) Kerr angle and ellipticity in the setup with layers (from left to right) of semi-infinite sapphire, monolayer CrBr3, and semi-infinite vacuum. (f) Kerr angle and ellipticity in the setup with layers (from left to right) of semi-infinite fused silica, monolayer CrBr3, and semi-infinite vacuum. A monolayer thickness $d = 6.07$ Å is used for monolayer CrBr3.